\documentclass[9pt]{elife}
\usepackage{epsfig,graphics,subfigure,amsmath,subeqnarray,color,xcolor, float}

\usepackage{lipsum} 
\usepackage[version=4]{mhchem}
\usepackage{siunitx}
\usepackage{array,longtable}
\DeclareSIUnit\Molar{M}
\newcommand{\refstyle}[1]{{\itshape\bfseries\color{eLifeMediumGrey}#1}}
\title{Swimming eukaryotic microorganisms exhibit a
universal speed distribution}

\author[1,2\authfn{1}*]{Maciej Lisicki}
\author[1\authfn{1}]{Marcos F. Velho Rodrigues}
\author[1]{Raymond E. Goldstein}
\author[1*]{Eric Lauga}
\affil[1]{Department of Applied Mathematics and Theoretical Physics, University of Cambridge, CB3 0WA Cambridge, United Kingdom}
\affil[2]{Institute of Theoretical Physics, Faculty of Physics, University of Warsaw, Warsaw, Poland}

\corr{maciej.lisicki@fuw.edu.pl}{ML}
\corr{e.lauga@damtp.cam.ac.uk}{EL}

\contrib[\authfn{1}]{These authors contributed equally to this work.}

\begin{document}

\maketitle

\begin{abstract}
One approach to quantifying biological diversity consists of characterizing the 
statistical distribution of specific properties of a taxonomic group or habitat. 
Microorganisms living in fluid environments, and for whom motility is key, 
exploit propulsion resulting from a rich variety of shapes, forms, and swimming 
strategies.  Here, we explore the variability of swimming speed for 
unicellular eukaryotes based on published data.  The data naturally partitions into that
from flagellates (with a small number of flagella) and from ciliates (with tens or 
more).  Despite the morphological and size differences between these groups, 
each of the two probability distributions of swimming speed are accurately 
represented by log-normal distributions, with good agreement holding even to 
fourth moments.  
Scaling of the distributions by a characteristic speed for each data set leads to
a collapse onto an apparently universal distribution.  These results suggest
a universal way for ecological niches to be populated by abundant microorganisms.
\end{abstract}

\section{Introduction}

Unicellular eukaryotes comprise a vast, diverse group of organisms that covers 
virtually all environments and habitats, displaying a menagerie of shapes and forms. 
Hundreds of species of the ciliate genus {\it Paramecium} \citep{wichterman1986} or 
flagellated {\it Euglena} \citep{buetow2011} are found in marine, brackish, and 
freshwater reservoirs; the green algae {\it Chlamydomonas} is distributed in soil and 
fresh water world-wide \citep{chlamy}; parasites from the genus {\it Giardia} colonize
intestines of several vertebrates \citep{Adam2001}.  One of the shared features of these organisms is their motility, crucial for nutrient acquisition and avoidance of danger \citep{bray2001}. In the process of evolution, single-celled organisms have developed in a variety of directions, and thus their rich morphology results in a large spectrum of swimming modes \citep{cappuccinelli1980}.

Many swimming eukaryotes actuate tail-like appendages called flagella or cilia in order to 
generate the required thrust \citep{sleigh1972}. This is achieved by actively generating deformations along the flagellum, giving rise to a complex waveform. The flagellar axoneme 
itself is a bundle of nine pairs of microtubule doublets surrounding two central microtubules, termed the "9+2" structure \citep{nicastro2005}, and cross-linking  
dynein motors, 
powered by ATP hydrolysis, perform mechanical work by promoting the relative sliding of filaments, resulting in bending deformations. 

\begin{figure}[t]
\centering{\includegraphics[width=100mm]{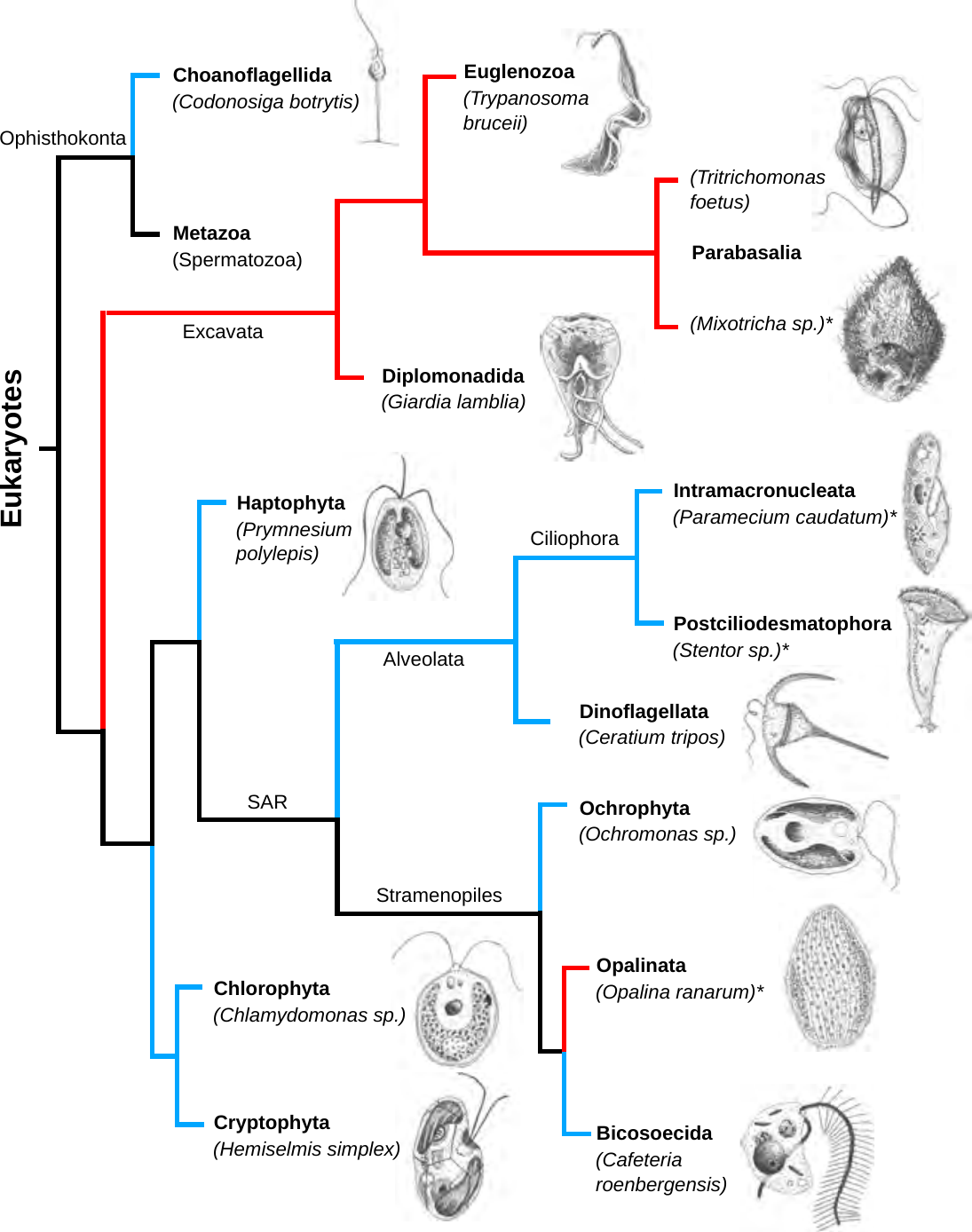}}
\caption{The tree of life (cladogram) for unicellular eukaryotes encompassing the phyla of organisms analyzed in the present study.  Aquatic organisms (living in marine, brackish, or freshwater environments) have their branches drawn in blue while parasitic organisms have their branches drawn in red. Ciliates are indicated by an asterisk after their names. For each phylum marked in bold font, a representative organism has been sketched next to its name. Phylogenetic data from \cite{opentree}.}
\label{fig:tree}
\end{figure}
 
Although eukaryotic flagella exhibit a diversity of forms and functions \citep{moran2014}, two large families, ``flagellates'' and ``ciliates'', can be distinguished 
by the shape and beating pattern of their flagella. 
Flagellates typically have a small number of long flagella distributed along the bodies, and they actuate them to generate thrust. The set of observed movement sequences includes planar undulatory waves and traveling helical waves, either from the base to the tip, or in the opposite direction \citep{jahn1972,brennen1977}. Flagella attached to the same body might follow different beating patterns, leading to a complex locomotion strategy that often relies also on the resistance the cell body poses to the fluid. 
In contrast, propulsion of ciliates derives from 
the motion of a layer of densely-packed and collectively-moving cilia, which are short hair-like flagella covering their bodies.  
The seminal review paper of \cite{brennen1977} lists a few examples from both groups, highlighting their shape, beat form, geometric characteristics and swimming properties. 
Cilia may also be used for transport of the surrounding fluid, and their cooperativity can lead to directed flow generation. In higher organisms this can be crucial for internal transport processes, as in cytoplasmic streaming within plant cells \citep{allen1978}, or the transport of ova from the ovary to the uterus in female mammals \citep{lyons2006}. 

Here, we turn our attention to these two morphologically different groups of swimmers to explore the variability of their propulsion dynamics within broad taxonomic groups. To this end, we have collected swimming speed data from literature for flagellated eukaryotes and ciliates and analyze them separately (we do not include spermatozoa since they lack (ironically) the capability to reproduce and are thus not living organisms; their
swimming characteristics have been studied by \cite{TamHosoi}). A careful examination of the statistical properties of the speed distributions for flagellates and ciliates shows that they are not only both captured by log-normal distributions but that, upon rescaling the data by a characteristic swimming speed for each data set, the speed distributions in both types of organisms are essentially identical.

\section{Results and Discussion}

We have collected swimming data on 189 unicellular eukaryotic microorganisms 
($N_f=112$ flagellates and $N_c=77$ ciliates) (see \refstyle{Appendix 1} 
and \refstyle{Appendix 1 - Source Data File 1}). \FIG{tree} shows a tree encompassing the phyla of organisms studied and sketches of a representative organism from each phylum. A large morphological variation is clearly visible. In addition, we delineate the branches involving aquatic organisms and parasitic species living within hosts. Both groups include ciliates and flagellates. 

\begin{figure}[h]
\includegraphics[width=140mm]{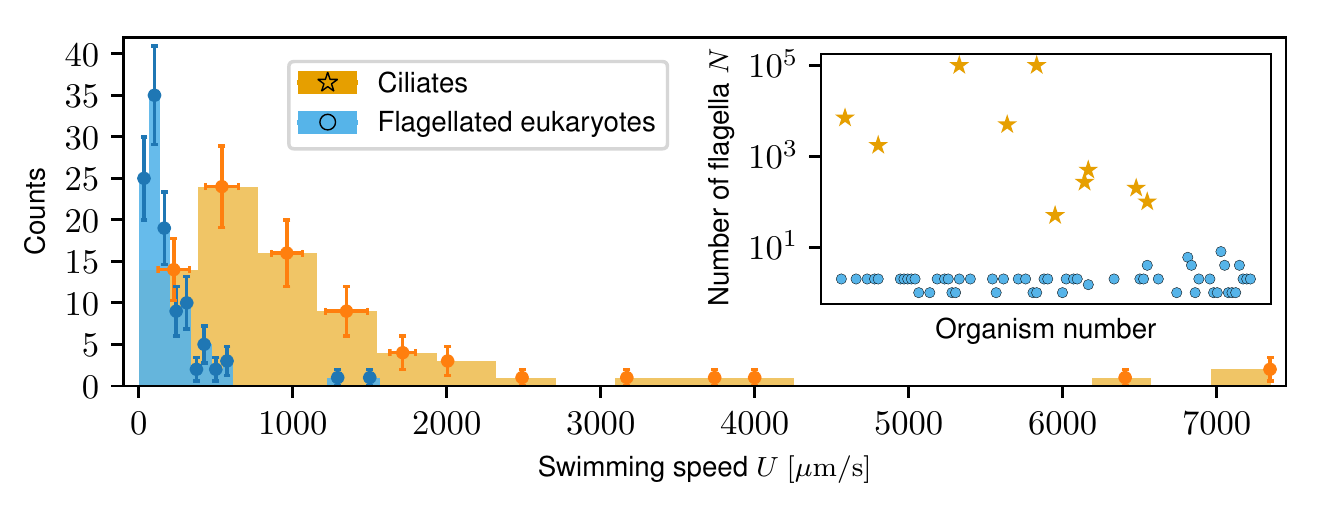}
\caption{Histograms of swimming speed for ciliates and flagellates demonstrate a similar character but different scales of velocities. Data points represent the mean and standard deviation of the data in each bin; horizontal error bars represent variability within each bin, 
vertical error bars show the standard deviation of the count. 
Inset: number of flagella displayed, where available, for each organism exhibits a clear morphological division between ciliates and flagellates.}
\label{fig:distribution}
\figsupp{Linear distribution of swimming speed data.  Symbols 
have been randomly placed vertically to avoid overlap.}{\includegraphics[width=\linewidth]{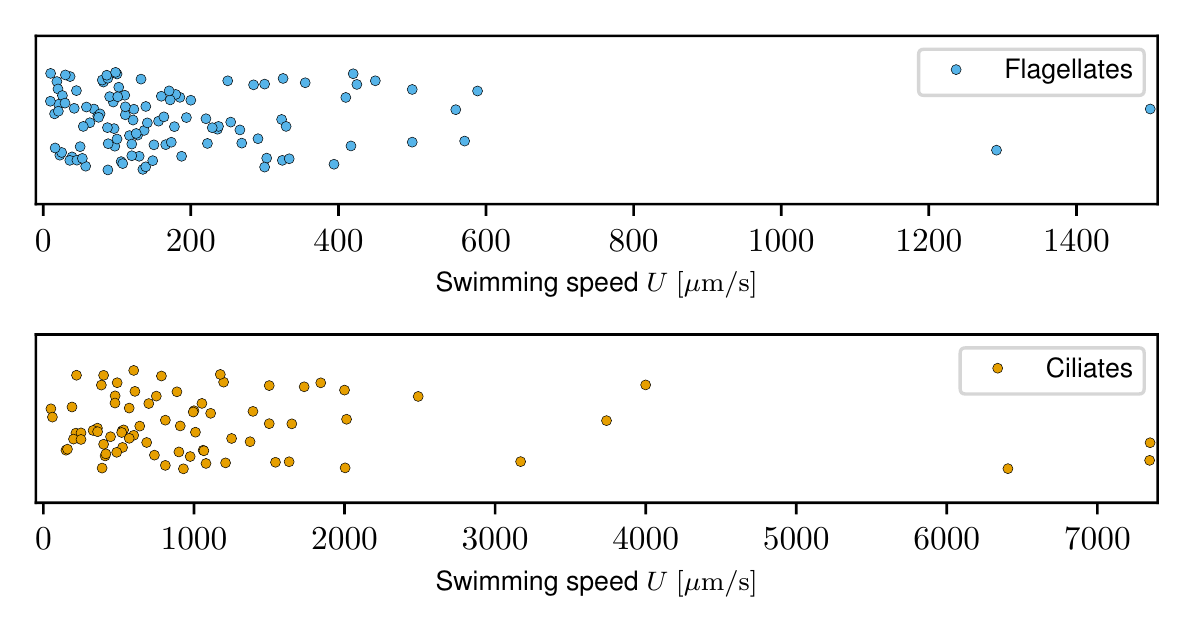}}
\label{figsupp:linear_distribution}
\figsupp{{Distribution of organism sizes in analyzed groups. Each histogram has been rescaled by the average cell size for each group. Although both distributions exhibit a qualitatively similar shape biased toward the low limit, no quantitative similarity is found.}}{\includegraphics[width=\linewidth]{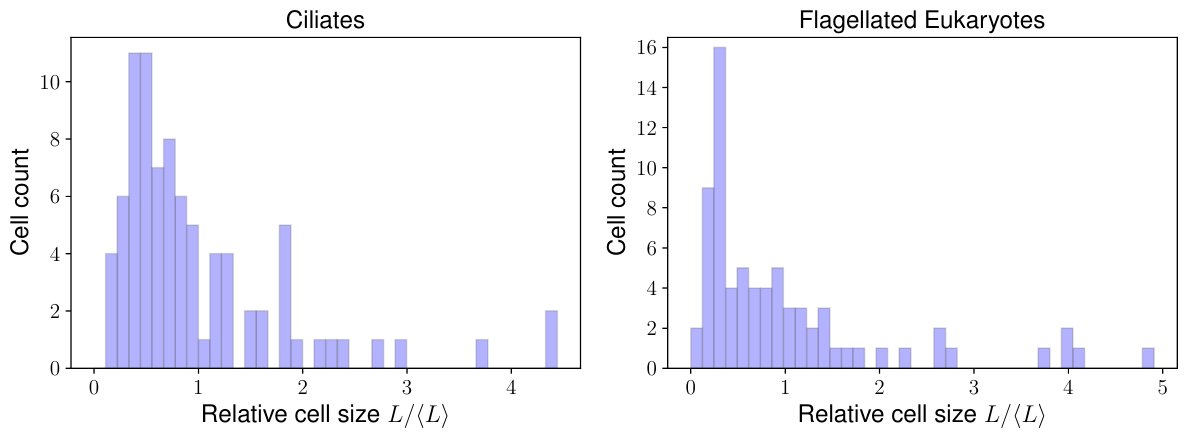}}
\label{figsupp:w_distribution}
\figsupp{{Distribution of Reynolds numbers for organisms in analyzed groups. Source data for the characteristic size $L$ and swimming speeds $U$ are listed in \refstyle{Appendix 1}.}}{\includegraphics[width=\linewidth]{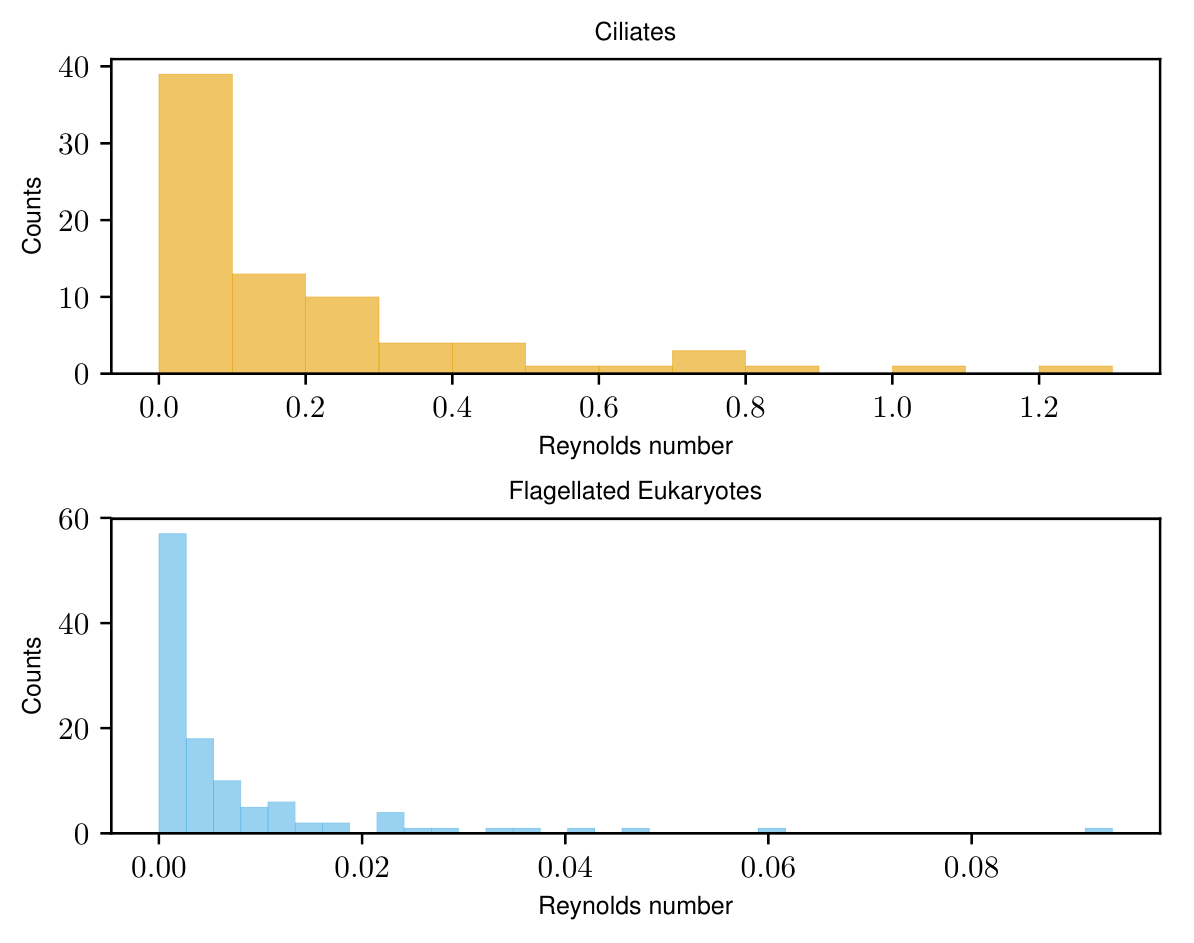}}
\label{figsupp:re_distribution}
\end{figure}

Due to the morphological and size differences between ciliates and flagellates, we investigate separately the statistical properties of each. \FIG{distribution} 
shows the two swimming speed histograms superimposed, based on
the raw distributions shown in \FIGSUPP[distribution]{linear_distribution},
where bin widths have been adjusted to their respective samples using the Freedman-Diaconis rule (see \refstyle{Materials and
Methods}).  Ciliates span a much larger range of speeds, up to 7 mm/s, whereas generally smaller flagellates remain in the sub-mm/s range. The inset shows that the number of flagella in both groups leads to a clear division. {To compare the two groups further, we have also collected information on the characteristic sizes of swimmers from the available literature, which we list in \refstyle{Appendix 1}. The average cell size differs fourfold between the populations (31 $\mu$m for flagellates and 132 $\mu$m for ciliates) and the distributions, plotted in \FIGSUPP[distribution]{w_distribution}, are biased towards the low-size end but they are quantitatively different. In order to explore the physical conditions, we used the data on sizes and speeds to compute the Reynolds number $\mathrm{Re}= UL/\nu$ for each organism, where $\nu=\eta/\rho$ is the kinematic viscosity of water, with $\eta$ the viscosity and $\rho$ the density. Since almost no data was available for the viscosity of the fluid in swimming speed measurements, we assumed the standard value $\nu=10^{-6}$ $\mu$m$^2/s$ for water for all organisms. The distribution of Reynolds numbers (\FIGSUPP[distribution]{re_distribution}), shows that ciliates and flagellates operate in different ranges of $\mathrm{Re}$, although for both groups $\mathrm{Re}<1$, imposing on them the same limitations of inertia-less Stokes flow \citep{LaLRN,lauga2009}. }

Furthermore, studies of green algae \citep{Short2006,Goldstein2015} 
show that an important distinction between the smaller, flagellated species
and the largest multicellular ones involves the relative importance of advection
and diffusion, as captured by the P{\'e}clet number 
$Pe=UL/D$, where $L$ is a typical organism size and $D$ is the 
diffusion constant of a relevant molecular species.  Using the average size $L$ of the 
cell body in each group of the present study ($L_{\rm fl}=31 \mu$m, $L_{\rm cil}=132$ $\mu$m) and the median
swimming speeds ($U_{\rm fl}=127$ $\mu$m/s, $U_{\rm cil}=784$ $\mu$m/s), and taking
$D=10^3$ $\mu$m$^2$/s, we find $Pe_{\rm fl}\sim 3.9$ and $Pe_{\rm cil}\sim 103$, 
which further justifies analyzing the groups separately;  they live
in different physical regimes.

\begin{figure}[t!]
\includegraphics[width=\linewidth]{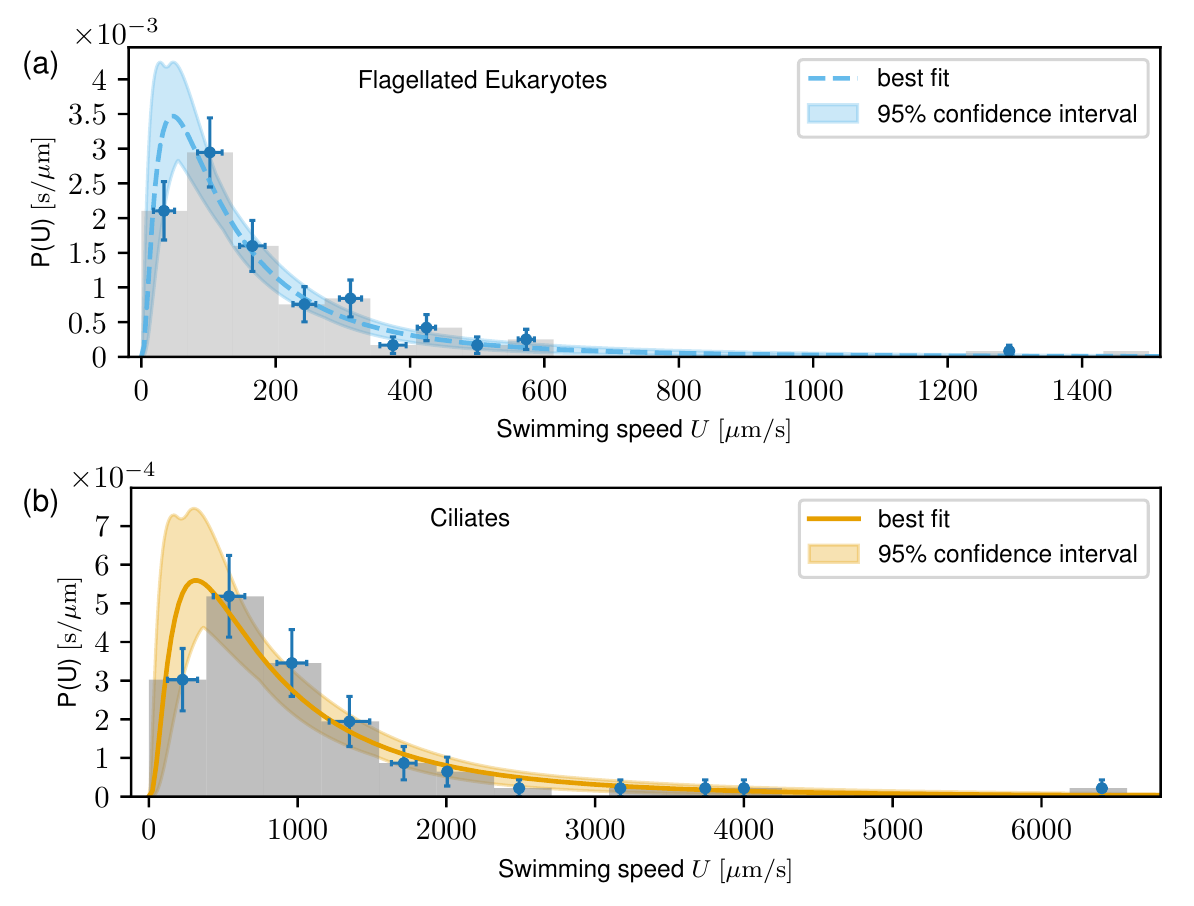}
\caption{Probability distribution functions of swimming speeds for flagellates (a) and ciliates (b) with the fitted log-normal distributions. Data points represent uncertainties as in \FIG{distribution}. Despite the markedly different scales of the distributions, they have similar shapes.}
\label{fig:lognorm_fits}
\figsupp{Higher moments of the swimming speed 
distributions {obtained from the data} compared with those {calculated from the fitted} log-normal distribution. The algebraic moments $\mathcal{M}_n$ are defined in Eq. (\ref{algmom}). {Error bars representing 95\% confidence intervals for fitted parameters, are obscured by  markers.}}{\includegraphics[width=\linewidth]{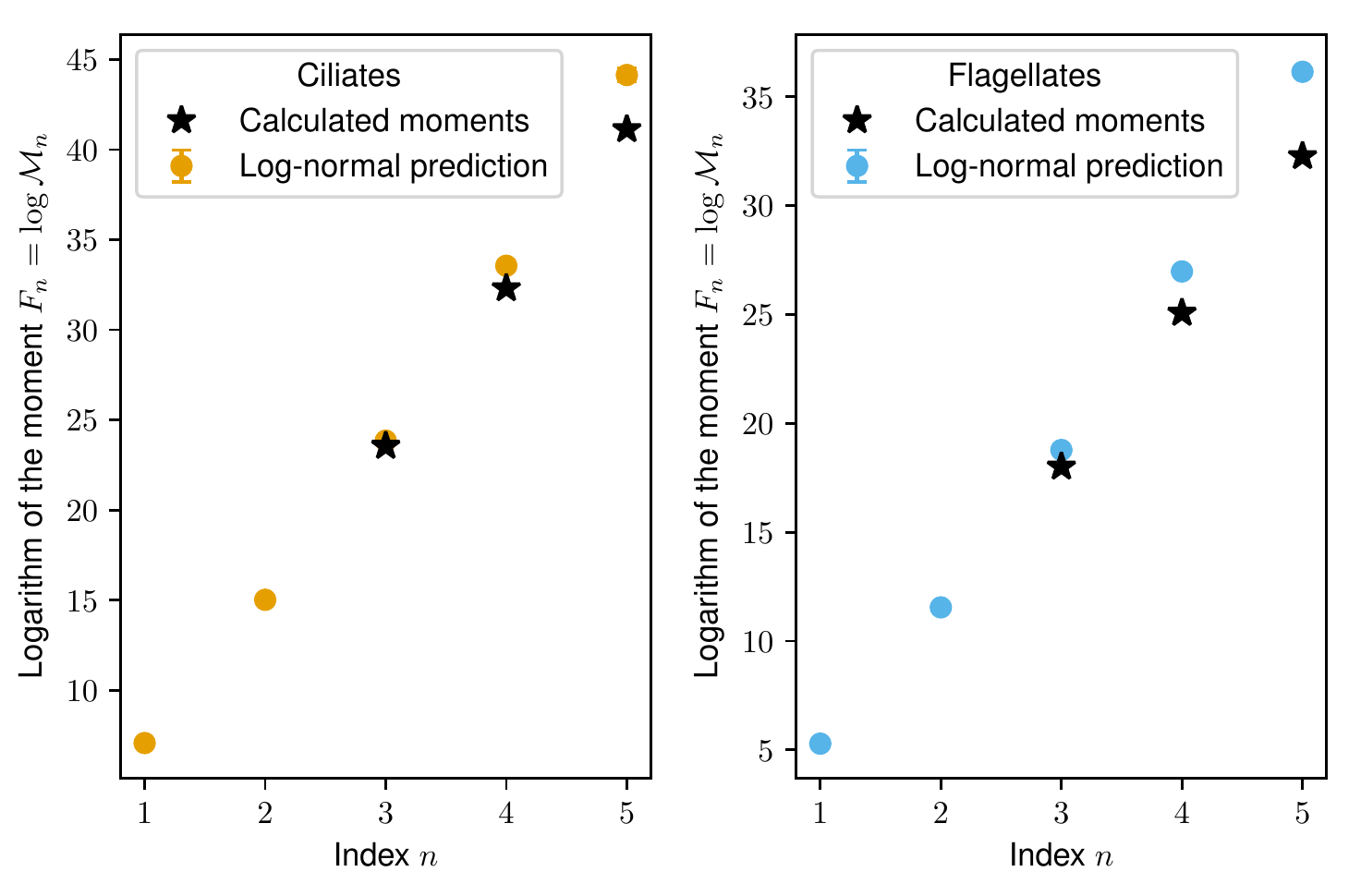}}
\label{figsupp:higher_moments}
\end{figure}

Examination of the mean, variance, kurtosis, and higher moments of the data sets suggest 
that the probabilities $P(U)$ of the swimming speed are well-described by log-normal distributions,
\begin{equation}
P(U) = \frac{1}{U \sigma \sqrt{2\pi} }\exp\left(-\frac{(\ln U - \mu)^2}{2\sigma^2}\right),	
\label{eq:lognormal}
\end{equation}
normalized as $\int_0^\infty dU P(U)=1$, 
where $\mu$ and $\sigma$ are the mean and the standard deviation of $\ln U$. The median $M$ of the distribution
is ${\rm e}^\mu$, with units of speed.
Log-normal distributions are widely observed across nature in areas such as ecology, physiology, geology and climate science, serving as an empirical model for complex processes shaping a system with many potentially interacting elements \citep{limpert2001}, particularly when the underlying processes involve proportionate fluctuations or multiplicative noise \citep{koch1966}. 

The results of fitting (see \refstyle{Materials and Methods}) are plotted in \FIG{lognorm_fits}, where the best fits are presented as solid curves, with the shaded areas representing 95\% confidence intervals. 
For flagellates, we find the $M_f = 127$ $\mu$m/s and $\sigma_f = 0.978$ while for ciliates, we obtain 
$M_c = 784$ $\mu$m/s and $\sigma_c = 0.936$.
Log-normal distributions are known to emerge from an (imperfect)  analogy to the Gaussian central limit theorem (see Materials and Methods). Since the data are accurately described by this distribution, we conclude that the published literature includes a sufficiently large amount of unbiased data to be able to see the whole distribution.   
  
\begin{figure}[t]
\includegraphics[width=\linewidth]{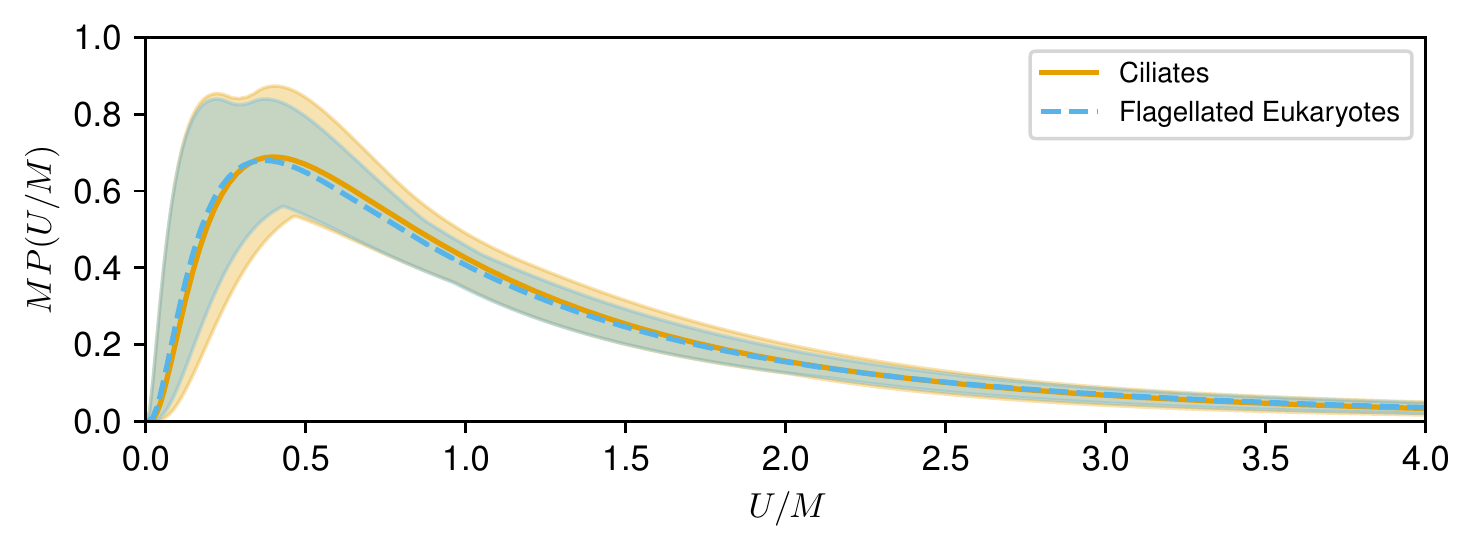}
\caption{Test of rescaling hypothesis.  Shown are the two fitted log-normal curves for flagellates and ciliates, each multiplied 
by the distribution median $M$, plotted versus speed normalized by $M$.  The distributions for show remarkable similarity and uncertainty of estimation.}
\label{fig:comparison}
\figsupp{Data collapse as in the main figure, but using the mean speeds $U^*$ instead of the median $M$.  A similar quality of 
data collapse is seen.}{\includegraphics[width=\linewidth]{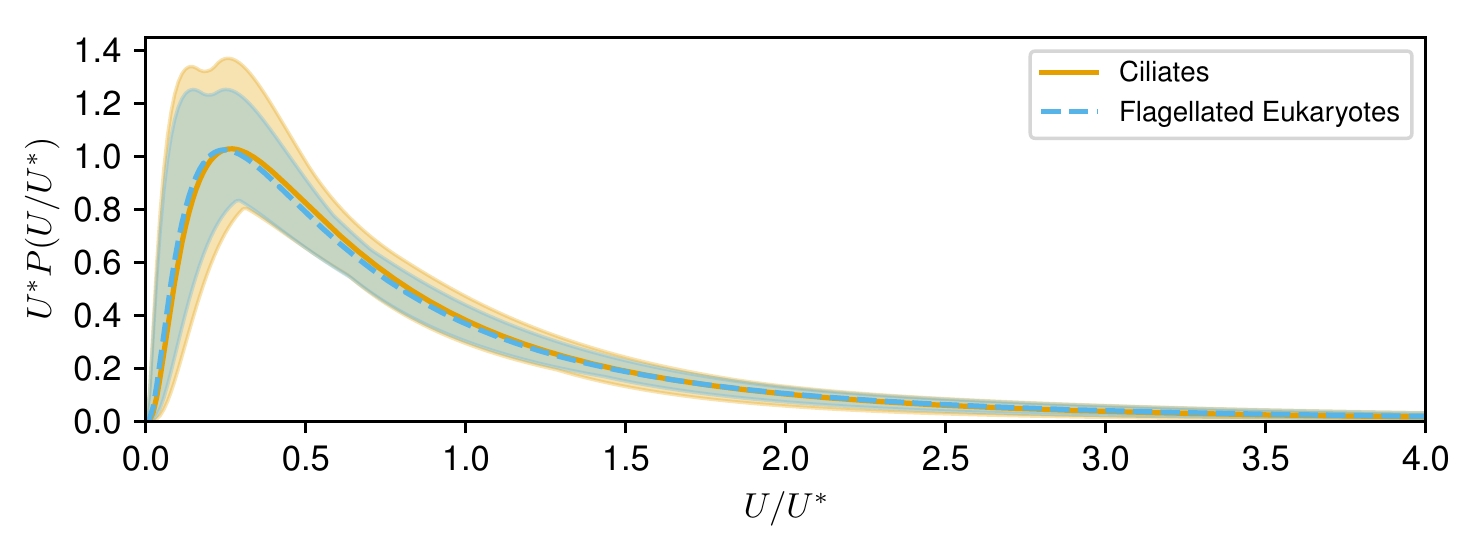}}
\label{figsupp:comparison_with_mean_speed}
\end{figure}

We next compare the statistical variability within groups by examining rescaled distributions \citep{Goldstein2018}.  As each has a characteristic speed $M$, we align the
peaks by plotting the distributions versus the variable $U/M$ for each group.  Since
$P$ has units of 1/speed, we are thus led to the form
$P(U,M)=M^{-1}F(U/M)$ for some function $F$.  For the
log-normal distribution, with $M$ the median, we find   
\begin{equation}
F(\xi)=\frac{1}{\xi \sigma \sqrt{2\pi} }\exp\left(-\frac{\ln^2 \xi}{2\sigma^2}\right)~,
\end{equation}
which now depends on the single parameter $\sigma$ and
has a median of unity by construction. To study the similarity of the two distributions we
plot the functions $F=MP(U/M)$ for each. As seen 
in \FIG{comparison}, the rescaled distributions are essentially indistinguishable, and this can be traced
back to the near identical values of the variances $\sigma$, which are within $5\%$ of each other.  The fitting uncertainties shown shaded in 
\FIG{comparison} suggest a very similar range of variability of the fitted distributions. Furthermore, 
both 
the integrated absolute difference between the 
distributions ($0.028$) and the Kullback-Leibler 
divergence ($0.0016$) are very small 
(see \refstyle{Materials and Methods}), demonstrating
the close similarity of the two distributions.  This similarity
is robust to the choice of characteristic speed, as shown in
\FIGSUPP[comparison]{comparison_with_mean_speed}, where the arithmetic
mean $U^*$ is used in place of the median.
 
In living cells, the sources for intrinsic variability within organisms are well characterized on the molecular and cellular level \citep{kirkwood2005} but less is known about variability within taxonomic groups. By dividing unicellular eukaryotes into two major groups on the basis of their difference in morphology, size and swimming strategy, we were able to capture in this paper the log-normal variability within each subset. Using a statistical analysis of the distributions as functions of the median swimming speed for each population we further found an almost identical distribution of swimming speeds for both types of organisms.   Our results suggest that the observed log-normal randomness captures a universal way for  ecological niches  to be populated by  abundant microorganisms with similar propulsion characteristics. {We note, however, that the distributions of swimming speeds among species do not necessarily reflect the distributions of swimming speeds among individuals, for which we have no available data.}

\section{Methods and Materials}
\label{methods}

\subsection{Data collection}

Data for ciliates were sourced from 26 research articles, 
while that for flagellates were extracted from 48 papers (see \refstyle{Appendix 1}). Notably, swimming speeds reported in the various studies have been measured under different physiological and environmental conditions, including temperature, viscosity, salinity, oxygenation, pH and light. 
Therefore we consider the data \textit{not} as representative of a uniform environment, but instead as arising from a random sampling of a wide range of environmental conditions. In cases where no explicit figure was given for $U$ in a paper, estimates were made using other available data where possible. Size of swimmers has also been included as a characteristic length for each organism. This, however, does not reflect the spread and diversity of sizes within populations of individual but is rather an indication of a typical size, as in the considered studies these data were not available. Information on anisotropy (different width/length) is also not included.

No explicit criteria were imposed for the inclusion in the analyses, apart from the biological classification (i.e. whether the organisms were unicellular eukaryotic ciliates/flagellates). We have used all the data found in literature for these organisms over the course of an extensive search. Since no selection was made, we believe that the observed statistical properties are representative for these groups.

\subsection{Data processing and fitting the log-normal distribution}

Bin widths in histograms in \FIG{distribution} and \FIG{lognorm_fits} have been chosen separately for ciliates and flagellated eukaryotes according to the Freedman-Diaconis rule \citep{freedman1981} taking into account the respective sample sizes and the spread of distributions. The bin width $b$ is then given by the number of observations $N$ and the interquartile range of the data $\mathrm{IQR}$ as
\begin{equation}
	b = 2 \frac{\mathrm{IQR}}{N^{1/3}}.
\end{equation}

Within each bin in \FIG{lognorm_fits}, we calculate the mean and the standard deviation for the binned data, which constitute the horizontal error bars. The vertical error bars reflect the uncertainty in the number of counts $N_j$ in bin $j$. This is estimated to be Poissonian, and thus the absolute error amounts to $\sqrt{N_j}$. Notably, the relative error decays with the number of counts as $1/\sqrt{N_j}$.

In fitting the data, we employ the log-normal distribution Eq.
\eqref{eq:lognormal}.  In general, from from data comprising 
$N$ measurements, labelled $x_i$ ($i=1,...,N$), the $n$-th arithmetic moment ${\cal M}_n$ is the expectation $\mathbb{E}(X^n)$, or
\begin{equation}\label{algmom}
	{\cal M}_n = \frac{1}{N}\sum_{i=1}^N x_i^n
\end{equation}
Medians of the data were found by sorting the list of values and picking the middlemost value.
For a log-normal distribution, the arithmetic moments are given solely by $\mu$ and $\sigma$ of the associated normal distribution as
\begin{equation}\label{algebr}
	 {\cal M}_n = M^n\Sigma^{n^2},
\end{equation}
where we have defined $M=\exp(\mu)$ and $\Sigma=\exp(\sigma^2/2)$, 
and note that $M$ is the median of the 
distribution.  Thus, the mean is $M\Sigma$ 
and the variance is $M^2\Sigma^2\left(\Sigma^2-1\right)$.  
From the first and second moments, we estimate 
\begin{equation}
	\mu = \ln \left( \frac{{\cal M}_1^2}{\sqrt{{\cal M}_2}}\right) \ \ \ \ {\rm and}
    \ \ \ \ 
	\sigma^2 = \ln\left( \frac{{\cal M}_2}{{\cal M}_1^2}\right).
\end{equation}
Having estimated $\mu$ and $\sigma$, we can compute the higher order moments from Eq. \eqref{algebr} 
and compare to those calculated directly from the data, as shown in \FIGSUPP[lognorm_fits]{higher_moments}. 

To fit the data, we have used both the MATLAB fitting routines and the Python \texttt{scipy.stats} module. From these fits we estimated the shape and scale parameters and the 95\% confidence intervals in \FIG{lognorm_fits} and \FIG{comparison}. We emphasize that the fitting procedures use the raw data via the maximum likelihood estimation method, and not the processed histograms, hence the estimated parameters are insensitive to the binning procedure. 

For rescaled distributions, the average velocity for each group of organisms was calculated as $U^\ast = \frac{1}{N_i} \sum_{i=1}^{N_i} U_i$, with $i\in\{c,f\}$. Then, data in each subset have been rescaled by the area under the fitted curve to ensure that the resulting probability density functions $p_i$ are normalized as
\begin{equation}
	\int_{0}^{\infty} p_i(x)\mathrm{d}x = 1.
\end{equation}

In characterizations of biological or ecological diversity, it is often assumed that the examined variables are Gaussian, and thus the distribution of many uncorrelated variables attains the normal distribution by virtue of the Central Limit Theorem (CLT). In the case when random variables in question are positive and have a log-normal distribution, no analogous explicit analytic result is available.  Despite that, there is general agreement that a sum of independent log-normal random variables can be well approximated by another log-normal random variable. It has been proven by \cite{szyszkowicz2009} that the sum of identically distributed equally and positively correlated joint log-normal distributions converges to a log-normal distribution of known characteristics but for uncorrelated variables only estimations are available \citep{beaulieu1995}. We use these results to conclude that our distributions contain enough data to be unbiased and seen in full.

\subsection{Comparisons of distributions}

In order to quantify the differences between the fitted distributions, we define the integrated absolute difference $\Delta$ between two probability distributions $p(x)$ and $q(x)$ ($x>0$) as
\begin{equation}
	\Delta = \int_{0}^{\infty} |p(x)-q(x)|\mathrm{d}x.
\end{equation}
As the probability distributions are normalized, this is a measure of their relative 'distance'. As a second measure, we use the Kullback-Leibler 
divergence \citep{kullback1951},
\begin{equation}
	D(p,q) =  \int_{0}^{\infty} p(x)\ln\left(\frac{p(x)}{q(x)}\right)\mathrm{d}x.
\end{equation}
Note that $D(p,q)\neq D(q,p)$ and therefore $D$ is not a distance metric in the space of probability distributions.

\section{Acknowledgments}

This project has received funding from the European Research Council (ERC) under the European Union’s Horizon 2020 research and innovation program (grant agreement 682754 to EL),
and from Established Career Fellowship EP/M017982/1
from the Engineering and Physical Sciences Research Council and Grant 7523 from the Gordon and Betty Moore 
Foundation (REG).



 \appendix
 \begin{appendixbox}
 \label{first:app}
 The Appendix contains the data which form the basis of our study. The tables contain data on the sizes and swimming speed of ciliates organisms and  flagellated eukaryotes from the existing literature. Data for ciliates were sourced from 26 research articles, while data for the flagellates were extracted from 48 papers.  In the cases where two or more sources reported contrasting figures for the swimming speed, the average value is reported in our tables.  The data itself is available in \refstyle{Appendix 1 - Source Data File 1}.

 \section{Data for swimming flagellates}

 Abbreviations:\\
dflg. – dinoflagellata; dph – dinophyceae; chlph. – chlorophyta; ochph. (het.) – ochrophyta (heterokont); srcm. – sarcomastigophora, pyr. – pyramimonadophyceae; prym. – prymnesiophyceae; dict. – dictyochophyceae; crypt. – cryptophyceae; chrys. – chrysophyceae 
 {\scriptsize
\begin{longtable}{|p{3cm}|p{1cm}p{1.7cm}|p{1.05cm}p{1.05cm}|p{3.2cm}|}
\hline
 \textbf{Species}&\textbf{Phylum}&\textbf{Class}&$L$[$\mu\mathrm{m}$] &$U$ [$\mu\mathrm{m}/s$]&\textbf{References} \\ \hline
\textit{Alexandrium minutum}&dflg.&dph.&21.7 &222.5& \cite{L270}\\
\textit{Alexandrium ostenfeldii}&dflg.&dph.&41.1 &110.5& \cite{L270}\\
\textit{Alexandrium tamarense}&dflg.&dph.&26.7 &200& \cite{L270}\\
\textit{Amphidinium britannicum}&dflg.&dph.&51.2 &68.7& \cite{L154}\\
\textit{Amphidinium carterae}&dflg.&dph.&16 &81.55& \cite{83,L154}\\
\textit{Amphidinium klebsi}&dflg.&dph.&35 &73.9& \cite{83}\\
\textit{Apedinella spinifera }&ochph. (het.)&dict.&8.25 &132.5&\cite{L160}\\
\textit{Bodo designis }&euglenozoa&kinetoplastea&5.5 & 39& \cite{L180}\\
\textit{Brachiomonas submarina}&chlph.&chlorophyceae&27.5 & 96& \cite{L154}\\
\textit{Cachonina (Heterocapsa) niei}&dflg.&dph.&21.4 &302.8& \cite{L158,L163}\\
\textit{Cafeteria roenbergensis}&bygira  (heterokont)&bicosoecida&2 &94.9& \cite{L176}\\
\textit{Ceratium cornutum}&dflg.&dph.&122.3 &177.75& \cite{L158,L233}\\
\textit{Ceratium furca}&dflg.&dph.&122.5 &194& \cite{163}\\
\textit{Ceratium fusus}&dflg.&dph.&307.5 &156.25& \cite{163}\\
\textit{Ceratium hirundinella}&dflg.&dph.&397.5 &236.1& \cite{L158}\\
\textit{Ceratium horridum}&dflg.&dph.&225 &20.8& \cite{163}\\
\textit{Ceratium lineatus}&dflg.&dph.&82.1 &36& \cite{L179}\\
\textit{Ceratium longipes}&dflg.&dph.&210 &166& \cite{163}\\
\textit{Ceratium macroceros}&dflg.&dph.&50 &15.4& \cite{163}\\
\textit{Ceratium tripos}&dflg.&dph.&152.3 &121.7&\cite{163,L154}\\
\textit{Chilomonas paramecium }&cryptophyta&crypt.&30 &111.25&\cite{137,116,83}\\
\textit{Chlamydomonas reinhardtii}&chlph.&chlorophyceae&10 &130& \cite{83,L161,L203}\\
\textit{Chlamydomonas moewusii}&chlph.&chlorophyceae&12.5 &128& \cite{83}\\
\textit{Chlamydomonas sp. }&chlph.&chlorophyceae&13 &63.2& \cite{143,L96,L154}\\
\textit{Crithidia deanei}&euglenozoa&kinetoplastea&7.4 &45.6& \cite{L170}\\
\textit{Crithidia fasciculata}&euglenozoa&kinetoplastea&11.1 &54.3& \cite{L170}\\
\textit{Crithidia \linebreak (Strigomonas) oncopelti}&euglenozoa&kinetoplastea&8
.1&18.5& \cite{L161,83}\\
\textit{Crypthecodinium cohnii}&dflg.&dph.&n/a &122.8& \cite{L179}\\
\textit{Dinophysis acuta}&dflg.&dph.&65 &500& \cite{163}\\
\textit{Dinophysis ovum}&dflg.&dph.&45 &160& \cite{L144}\\
\textit{Dunaliella} sp. &chlph.&chlorophyceae&10.8 &173.5& \cite{83,L154}\\
\textit{Euglena gracilis}&euglenozoa&euglenida   (eugl.)&47.5 &111.25& \cite{137,116,83}\\
\textit{Euglena viridis}&euglenozoa&euglenida  (eugl.)&58 &80& \cite{108,L161,L96}\\
\textit{Eutreptiella gymnastica}&euglenozoa&euglenida (aphagea)&23.5 &237.5& \cite{L160}\\
\textit{Eutreptiella } sp. R&euglenozoa&euglenida&50 &135&\cite{L160}\\
\textit{Exuviaella baltica \linebreak(Prorocentrum balticum)}&dflg.&dph.&15.5 &138.9& \cite{L162}\\
\textit{Giardia lamblia}&srcm. &zoomastigophora&11.25 &26& \cite{N11,M16,M17}\\
\textit{Gonyaulax polyedra}&dflg.&dph.&39.2 &254.05& \cite{97,83,L64}\\
\textit{Gonyaulax polygramma}&dflg.&dph.&46.2 &500& \cite{L158}\\
\textit{Gymnodinium aureolum}&dflg.&dph.&n/a &394& \cite{L173}\\
\textit{Gymnodinium \linebreak sanguineum (splendens)}&dflg.&dph.&47.6 &220.5& \cite{L64,L158}\\
\textit{Gymnodinium simplex}&dflg.&dph.&10.6 &559& \cite{L145}\\
\textit{Gyrodinium aureolum}&dflg.&dph.&30.5 &139& \cite{L154,L160}\\
\textit{Gyrodinium dorsum} \linebreak(bi-flagellated)&dflg.&dph.&37.5 &324& \cite{97,83,L64,L158, BandW}\\
\textit{Gyrodinium dorsum} \linebreak(uni-flagellated)&dflg.&dph.&34.5 &148.35& \cite{99}\\
\textit{Hemidinium nasutum}&dflg.&dph.&27.2 &105.6& \cite{L158,L233}\\
\textit{Hemiselmis simplex}&cryptophyta&crypt.&5.25 &325& \cite{L160}\\
\textit{Heterocapsa pygmea}&dflg.&dph.&13.5 &102.35& \cite{L154}\\
\textit{Heterocapsa rotundata}&dflg.&dph.&12.5 &323&\cite{L145}\\
\textit{Heterocapsa triquetra}&dflg.&dph.&17 &97& \cite{L180}\\
\textit{Heteromastix pyriformis }&chlph.&nephrophyseae&6 &87.5& \cite{L160}\\
\textit{Hymenomonas carterae }&haptophyta&prym.&12.5 &87& \cite{L154}\\
\textit{Katodinium rotundatum (Heterocapsa rotundata)}&dflg.&dph.&10.8 &425& \cite{L158,L160}\\
\textit{Leishmania major}&euglenozoa&kinetoplastea&12.5 &36.4& \cite{L170}\\
\textit{Menoidium cultellus}&euglenozoa&euglenida (eugl.)&45 &136.75& \cite{108,221}\\
\textit{Menoidium incurvum}&euglenozoa&euglenida (eugl.)&25 &50& \cite{L96,83}\\
\textit{Micromonas pusilla}&chlph.&mamiellophyceae&2 &58.5& \cite{L154,L160}\\
\textit{Monas stigmata}&ochph. (het.)&chrys.&6 &269& \cite{83}\\
\textit{Monostroma angicava}&chlph.&ulvophyceae&6.7 &170.55& \cite{M20}\\
\textit{Nephroselmis pyriformis }&chlph.&nephrophyseae&4.8 &163.5& \cite{L154}\\
\textit{Oblea rotunda}&dflg.&dph.&20 &420&\cite{L144}\\
\textit{Ochromonas danica}&ochph. (het.)&chrys.&8.7 &77& \cite{L267}\\    
\textit{Ochromonas malhamensis}&ochph. (het.)&chrys.&3 &57.5& \cite{107}\\
\textit{Ochromonas minima}&ochph. (het.)&chrys.&5 &75& \cite{L160}\\
\textit{Olisthodiscus luteus}&ochph. (het.)&raphidophyceae&22.5 &90& \cite{L154,L160}\\
\textit{Oxyrrhis marina}&dflg.&oxyrrhea&39.5 &300&\cite{L143,L179}\\
\textit{Paragymnodinium shiwhaense}&dflg.&dph.&10.9 &571& \cite{L173}\\
\textit{Paraphysomonas vestita}&ochph. (het.)&chrys.&14.7 &116.85& \cite{L177}\\
\textit{Pavlova lutheri }&haptophyta&pavlovophyceae&6.5 &126& \cite{L154}\\
\textit{Peranema trichophorum}&euglenozoa&euglenida  (heteronematales)&45 &20& \cite{L96,83,BandW}\\
\textit{Peridinium bipes}&dflg.&dph.&42.9 &291& \cite{L179}\\
\textit{Peridinium cf. quinquecorne}&dflg.&dph.&19 &1500& \cite{L154,L158,L169}\\
\textit{Peridinium cinctum}&dflg.&dph.&47.5 &120& \cite{L154,L158,L233}\\
\textit{Peridinium (Protoperidinium) claudicans} &dflg.&dph.&77.5 &229& \cite{163}\\
\textit{Peridinium (Protoperidinium) crassipes} &dflg.&dph.&102 &100& \cite{163}\\
\textit{Peridinium foliaceum}&dflg.&dph.&30.6 &185.2& \cite{L64}\\
\textit{Peridinium (Bysmatrum) gregarium }&dflg.&dph.&32.5 &1291.7& \cite{L158}\\
\textit{Peridinium (Protoperidinium) ovatum }&dflg.&dph.&61 &187.5&\cite{163}\\
\textit{Peridinium (Peridiniopsis) penardii }&dflg.&dph.&28.8 &417&\cite{L156}\\
\textit{Peridinium (Protoperidinium) pentagonum }&dflg.&dph.&92.5 &266.5&\cite{163}\\
\textit{Peridinium (Protoperidinium) subinerme }&dflg.&dph.&50 &285& \cite{163}\\
\textit{Peridinium trochoideum}&dflg.&dph.&25 &53& \cite{L158}\\
\textit{Peridinium umbonatum}&dflg.&dph.&30 &250& \cite{L158,L233}\\
\textit{Phaeocystis pouchetii}&haptophyta&prym.&6.3 &88& \cite{L154}\\
\textit{Polytoma uvella }&chlph.&chlorophyceae&22.5 &100.9& \cite{143,83,L96}\\
\textit{Polytomella agilis }&chlph.&chlorophyceae&12.4 &150& \cite{84,85,83,L161}\\
\textit{Prorocentrum mariae-lebouriae}&dflg.&dph.&14.8 &141.05& \cite{L64,L154,L159}\\
\textit{Prorocentrum micans }&dflg.&dph.&45 &329.1& \cite{L154,L158}\\
\textit{Prorocentrum minimum}&dflg.&dph.&15.1 &107.7& \cite{L154,L159}\\
\textit{Prorocentrum redfieldii Bursa (P.triestinum)}&dflg.&dph.&33.2 &333.3& \cite{L155}\\
\textit{Protoperidinium depressum}&dflg.&dph.&132 &450& \cite{L144}\\
\textit{Protoperidinium granii (Ostf.) Balech}&dflg.&dph.&57.5 &86.1&\cite{L155}\\
\textit{Protoperidinium pacificum}&dflg.&dph.&54 &410& \cite{L144}\\
\textit{Prymnesium polylepis}&haptophyta&prym.&9.1 &45& \cite{M18}\\
\textit{Prymnesium parvum}&haptophyta&prym.&7.2 &30& \cite{M18}\\
\textit{Pseudopedinella pyriformis }&ochph. (het.)&dict.&6.5 &100& \cite{L160}\\
\textit{Pseudoscourfieldia marina }&chlph.&pyr.&4.1 &42& \cite{L154}\\
\textit{Pteridomonas danica }&ochph. (het.)&dict.&5.5 &179.45& \cite{L177}\\
\textit{Pyramimonas amylifera}&chlph.&pyr.&24.5 &22.5& \cite{L154}\\
\textit{Pyramimonas cf. disomata}&chlph.&pyr.&9 &355& \cite{L160}\\
\textit{Rhabdomonas spiralis}&euglenozoa&euglenida (aphagea)&27 &120&\cite{108}\\
\textit{Rhodomonas salina }&cryptophyta&crypt.&14.5 &588.5& \cite{L145,L173}\\
\textit{Scrippsiella trochoidea}&dflg.&dph.&25.3 &87.6& \cite{L64,L154,L155}\\
\textit{Spumella } sp.&ochph. (het.)&chrys.&10 &25& \cite{L180}\\
\textit{Teleaulax } sp.&cryptophyta&crypt.&13.5 &98& \cite{L173}\\
\textit{Trypanosoma brucei}&euglenozoa&kinetoplastea&18.8 &20.5& \cite{L193}\\
\textit{Trypanosoma cruzi}&euglenozoa&kinetoplastea&20 &172& \cite{L133,BandW}\\
\textit{Trypanosoma vivax}&euglenozoa&kinetoplastea&23.5 &29.5& \cite{M14}\\
\textit{Trypanosoma evansi}&euglenozoa&kinetoplastea&21.5 &16.1& \cite{M14}\\
\textit{Trypanosoma congolense}&euglenozoa&kinetoplastea&18 &9.7& \cite{M14}\\
\textit{Tetraflagellochloris mauritanica}&chlph.&chlorophyceae&4 &300& \cite{M15}\\
\hline
\end{longtable}}

  \section{Data for swimming ciliates}
 
   Abbreviations:\\
 imnc. = intramacronucleata; pcdph. = postciliodesmatophora; olig. – oligohymenophorea; spir. – spirotrichea; hettr. – heterotrichea; lit. – litostomatea; eugl. – euglenophyceae
 
 {\scriptsize
\begin{longtable}{|p{2.8cm}|p{1cm}p{1.3cm}|p{1.1cm}p{1.1cm}|p{3.2cm}|}
\hline
 \textbf{Species}&\textbf{Phylum}&\textbf{Class}&$L$ [$\mu\mathrm{m}$]&$U$ [$\mu\mathrm{m}/s$]&\textbf{References} \\ \hline
\textit{Amphileptus gigas}&imnc.&lit.&808 & 608& \cite{47}\\
\textit{Amphorides quadrilineata}&imnc.&spir.&138 &490& \cite{L144}\\
\textit{Balanion comatum}&imnc.&prostomatea&16&220& \cite{L180}\\
\textit{Blepharisma}&pcdph.& hettr.&350 &600& \cite{L157,L161}\\
\textit{Coleps hirtus}&imnc.&prostomatea&94.5 &686& \cite{47}\\
\textit{Coleps} sp.&imnc.&prostomatea&78 &523&\cite{47}\\
\textit{Colpidium striatum}&imnc.&olig.&77 &570& \cite{L264}\\
\textit{Condylostoma patens}&pcdph.& hettr.&371 &1061& \cite{47,149}\\
\textit{Didinium nasutum}&imnc.&lit.&140 &1732& \cite{47,149,L161,L157}\\
\textit{Euplotes charon}&imnc.&spir.&66 &1053& \cite{47}\\
\textit{Euplotes patella}&imnc.&spir.&202 &1250& \cite{47}\\
\textit{Euplotes vannus}&imnc.&spir.&82 &446& \cite{M40,M41}\\
\textit{Eutintinnus cf. pinguis}&imnc.&spir.&147 &410& \cite{L144}\\
\textit{Fabrea salina}&pcdph.& hettr.&184.1 &216& \cite{L271}\\
\textit{Favella panamensis}&imnc.&spir.&238 &600& \cite{L144}\\
\textit{Favella} sp.&imnc.&spir.&150 &1080& \cite{L144}\\
\textit{Frontonia} sp.&imnc.&olig.&378.5 &1632& \cite{47}\\
\textit{Halteria grandinella}&imnc.&spir.&50 &533& \cite{47,L236}\\
\textit{Kerona polyporum}&imnc.&spir.&107 &476.5& \cite{47}\\
\textit{Laboea strobila}&imnc.&spir.&100 &810& \cite{L144}\\
\textit{Lacrymaria lagenula}&imnc.&lit.&45 &909& \cite{47}\\
\textit{Lembadion bullinum}&imnc.&olig.&43 &415& \cite{47}\\
\textit{Lembus velifer}&imnc.&olig.&87 &200& \cite{47}\\
\textit{Mesodinium rubrum}&imnc.&lit.&38 &7350&\cite{L152,L209,L175}\\
\textit{Metopides contorta}&imnc.&armophorea&115 &359& \cite{47}\\
\textit{Nassula ambigua}&imnc.&nassophorea&143&2004& \cite{47}\\
\textit{Nassula ornata}&imnc.&nassophorea&282&750& \cite{47}\\
\textit{Opalina ranarum}& placidozoa (heterokont)&opalinea&350 &50& \cite{L113,L157}\\
\textit{Ophryoglena} sp.&imnc.&olig.&325 &4000& \cite{149}\\
\textit{Opisthonecta henneg}&imnc.&olig.&126 &1197& \cite{149,119}\\
\textit{Oxytricha bifara}&imnc.&spir.&282 &1210& \cite{47}\\
\textit{Oxytricha ferruginea}&imnc.&spir.&150 &400& \cite{47}\\
\textit{Oxytricha platystoma}&imnc.&spir.&130 &520& \cite{47}\\
\textit{Paramecium aurelia}&imnc.&olig.&244 &1650& \cite{47,48}\\
\textit{Paramecium bursaria}&imnc.&olig.&130 &1541.5& \cite{47,48}\\
\textit{Paramecium calkinsii}&imnc.&olig.&124 &1392& \cite{48,47}\\
\textit{Paramecium caudatum}&imnc.&olig.&225.5 &2489.35&\cite{48,L206}\\
\textit{Paramecium marinum}&imnc.&olig.&115 &930&\cite{47}\\
\textit{Paramecium multimicronucleatum}&imnc.&olig.&251 &3169.5& \cite{48}\\
\textit{Paramecium polycaryum}&imnc.&olig.&91 &1500& \cite{48}\\
\textit{Paramecium}  spp.&imnc.&olig.&200 &975& \cite{116,L157,L161}\\
\textit{Paramecium tetraurelia}&imnc.&olig.&124 &784& \cite{L200}\\
\textit{Paramecium woodruffi}&imnc.&olig.&160 &2013.5& \cite{48}\\
\textit{Porpostoma notatum}&imnc.&olig.&107.7 &1842.2& \cite{L176}\\
\textit{Prorodon teres}&imnc.&prostomatea&175&1066& \cite{47}\\
\textit{Spathidium spathula}&imnc.&lit.&204.5&526& \cite{47}\\
\textit{Spirostomum ambiguum}&pcdph.& hettr.&1045 &810& \cite{47}\\
\textit{Spirostomum} sp.&pcdph.& hettr.&1000 &1000& \cite{L157}\\
\textit{Spirostomum teres}&pcdph.& hettr.&450 &640& \cite{47}\\
\textit{Stenosemella steinii}&imnc.&spir.&83 &190& \cite{L144}\\
\textit{Stentor caeruleus}&pcdph.& hettr.&528.5 &1500& \cite{47}\\
\textit{Stentor polymorphus}&pcdph.& hettr.&208 &887& \cite{47,203,194}\\
\textit{Strobilidium spiralis}&imnc.&spir.&60 &330& \cite{L144}\\
\textit{Strobilidium velox}&imnc.&spir.&43 &150& \cite{L236}\\
\textit{Strombidinopsis acuminatum}&imnc.&spir.&80 &390& \cite{L144}\\
\textit{Strombidium claparedi}&imnc.&spir.&69.5 &3740& \cite{47}\\
\textit{Strombidium conicum}&imnc.&spir.&75 &570& \cite{L144}\\
\textit{Strombidium} sp.&imnc.&spir.&33 &360&\cite{L144}\\
\textit{Strombidium sulcatum}&imnc.&spir.&32.5 &995& \cite{L142,L176}\\
\textit{Stylonichia} sp.&imnc.&spir.&167 &737.5& \cite{47,149}\\
\textit{Tetrahymena pyriformis}&imnc.&olig.&72.8 &475.6& \cite{L157,L161,BandW}\\
\textit{Tetrahymena thermophila}&imnc.&olig.&46.7 &204.5& \cite{M39}\\
\textit{Tillina magna}&imnc.&colpodea&162.5&2000& \cite{47}\\
\textit{Tintinnopsis kofoidi}&imnc.&spir.&100 &400&\cite{L144}\\
\textit{Tintinnopsis minuta}&imnc.&spir.&40 &60& \cite{L144}\\
\textit{Tintinnopsis tubulosa}&imnc.&spir.&95 &160& \cite{L144}\\
\textit{Tintinnopsis vasculum}&imnc.&spir.&82 &250& \cite{L144}\\
\textit{Trachelocerca olor}&pcdph.&karyorelictea&267.5&900& \cite{47}\\
\textit{Trachelocerca tenuicollis}&pcdph.&karyorelictea &432&1111& \cite{47}\\
\textit{Uroleptus piscis}&imnc.&spir.&203 &487& \cite{47}\\
\textit{Uroleptus rattulus}&imnc.&spir.&400 &385& \cite{47}\\
\textit{Urocentrum turbo}&imnc.&olig.&90 &700& \cite{47}\\
\textit{Uronema filificum}&imnc.&olig.&25.7 &1372.7&\cite{L176}\\
\textit{Uronema marinum}&imnc.&olig.&56.9 &1010& \cite{L176}\\
\textit{Uronema sp.}&imnc.&olig.&25 &1175& \cite{L157,L161}\\
\textit{Uronychia transfuga}&imnc.&spir.&118 &6406&\cite{N14}\\
\textit{Uronychia setigera}&imnc.&spir.&64 &7347& \cite{N14}\\
\textit{Uronemella} spp.&imnc.&olig.&28 &250&\cite{M12}\\
\hline
\end{longtable}}

 \end{appendixbox}

\end{document}